# Set-Membership Constrained Conjugate Gradient Beamforming Algorithms

Lei Wang and Rodrigo C. de Lamare

*Abstract*—In this work a constrained adaptive filtering strategy based on conjugate gradient (CG) and set-membership (SM) techniques is presented for adaptive beamforming. A constraint on the magnitude of the array output is imposed to derive an adaptive algorithm that performs data-selective updates when calculating the beamformer's parameters. We consider a linearly constrained minimum variance (LCMV) optimization problem with the bounded constraint based on this strategy and propose a CG type algorithm for implementation. The proposed algorithm has data-selective updates, a variable forgetting factor and performs one iteration per update to reduce the computational complexity. The updated parameters construct a space of feasible solutions that enforce the constraints. We also introduce two time-varying bounding schemes to measure the quality of the parameters that could be included in the parameter space. A comprehensive complexity and performance analysis between the proposed and existing algorithms are provided. Simulations are performed to show the enhanced convergence and tracking performance of the proposed algorithm as compared to existing techniques.

## I. INTRODUCTION

Constrained adaptive filtering algorithms have been developed for several decades and found applications in many areas such as beamforming, system identification, and interference suppression in code-division multiple-access (CDMA) systems [1]-[6]. For example, in beamforming problems the constraint corresponds to the knowledge of the direction-of-arrival (DOA) of the desired user, which is assumed known beforehand by the receive antenna.

The popular constrained adaptive filtering algorithms include stochastic gradient (SG), recursive least squares (RLS) [1], affine projection (AP) [10], and conjugate gradient (CG) [11]. SG algorithms are attractive due to their low computational cost. However, the convergence of SG algorithms strongly depends on the selection of the step size value and thus they suffer from slow convergence or large misadjustment. RLS algorithms have fast convergence but are more complex to implement and may become unstable due to the divergence problem and numerical problems. The AP and CG algorithms can be viewed as alternatives with computational complexities and convergence performances between those of the SG and RLS algorithms. The AP algorithms require the inversion of a matrix whose dimension is given by the projection order [1], which can increase significantly the complexity of AP algorithms when the projection order is high. Moreover, AP algorithms have a convergence rate that is significantly faster than the SG algorithm and can be significantly slower than the RLS, depending on the step size and the environment. CG algorithms outperform AP algorithms in severe situations and have a comparable performance to the RLS, while requiring a lower complexity.

In certain scenarios where the adaptive filters have a large number of parameters, or operate under constraints on the power consumption, the use of standard constrained adaptive algorithms is limited due to the high computational complexity. In this context, reduced-rank algorithms can be a highly effective solution [7]-[24]. Furthermore, set-membership (SM) techniques have been introduced into adaptive filtering [25]-[31]. SM algorithms specify a bound on the magnitude of the estimation error or the array output, and can reduce the computational complexity due to data-selective updates. SM techniques usually rely on two steps: 1) information evaluation and 2) parameter update. If step 2) does not occur frequently, and step 1) does not require much complexity, the overall complexity can be reduced significantly. SM techniques with the SG, RLS, and AP algorithms have been reported in [28]-[32], in which the SM-based algorithms achieve computation reduction without performance degradation compared with conventional algorithms. A CG algorithm with data selective updates has been reported in [31], which is named data selective conjugate gradient (DS-CG) here.

The main contribution of this paper is to introduce an adaptive filtering strategy that combines the SM technique with the CG algorithm in order to solve a constrained optimization problem. The proposed algorithm, which is termed SM-CG, achieves a fast convergence that is close to that of the RLS algorithm, employs a variable forgetting factor that allows excellent convergence and tracking performance, and requires a computational complexity that is much lower than its conventional counterparts. We present a linearly constrained minimum variance (LCMV) optimization problem [33] related to an SM bounded constraint on the array output, and we perform the filter optimization using a CG technique. The parameters are updated only if the bounded constraint cannot be satisfied. For the update, we introduce a CG-based vector $v \in \mathbb{C}^{m \times 1}$ to create a relation with the received covariance matrix $R = \mathbb{E}[rr^H]$ and the array response of the desired user $a(\theta_0)$, namely, $v = R^{-1}a(\theta_0)$, where $m$ denotes the filter length, $r \in \mathbb{C}^{m \times 1}$ is the received vector, and $\theta_0$ is the DOA of the desired user. This procedure avoids a matrix inversion associated with the constraint on the array response. Unlike the work reported in [31], the proposed SM-CG method employs a variable forgetting factor to improve

Communications Research Group, Department of Electronics, University of York, York YO10 5DD, UK. Email:{lw517,rcdl500}@ohm.york.ac.uk

the convergence and tracking performance, a strategy based on an auxiliary vector to impose the constraint on the array response that makes our algorithm computationally simpler than [31] and a time-varying bound. We also introduce two time-varying bounds that extend the work in [37] to beamforming and incorporate parameter and interference dependence to reflect the characteristics of the environment for improving the tracking performance of the proposed algorithm in dynamic scenarios. Compared with the fixed bounded constraint in the existing algorithm, the time-varying bounding schemes are very effective at reducing the update rate and enhancing the output performance in dynamic scenarios. A complexity analysis is provided to show the advantage of the proposed SM-CG algorithm over the existing methods. To evaluate the proposed algorithm, we consider a beamforming application and the output signal-to-interference-plus-noise ratio (SINR) performance. We compare the proposed algorithm with the existing algorithms under both stationary and non-stationary scenarios.

The paper is structured as follows. Section II briefly describes the system model and states the LCMV optimization problem. Section III describes the constrained SM adaptive filtering framework, whereas Section IV presents the proposed constrained SM-CG adaptive filtering algorithm. Section V introduces two strategies to compute time-varying bounds. Section VI presents and discusses the simulation results, while Section VII gives the concluding remarks.

## II. System Model and LCMV Optimization Problem

Let us suppose that $q$ narrowband signals impinge on a uniform linear array (ULA) of $m$ ($q \leq m$) sensor elements. The sources are assumed to be in the far field with DOAs $\theta_0, \ldots, \theta_{q-1}$. The received vector $\boldsymbol{r}$ can be modeled as

$$\boldsymbol{r} = \boldsymbol{A}(\boldsymbol{\theta})\boldsymbol{s} + \boldsymbol{n}, \tag{1}$$

where $\boldsymbol{\theta} = [\theta_0, \ldots, \theta_{q-1}]^T \in \mathbb{R}^{q \times 1}$ is the DOAs, $\boldsymbol{A}(\boldsymbol{\theta}) = [\boldsymbol{a}(\theta_0), \ldots, \boldsymbol{a}(\theta_{q-1})] \in \mathbb{C}^{m \times q}$ contains the steering vectors $\boldsymbol{a}(\theta_k) = [1, e^{-2\pi j \frac{d}{\lambda_c} cos\theta_k}, \ldots, e^{-2\pi j (m-1) \frac{d}{\lambda_c} cos\theta_k}]^T \in \mathbb{C}^{m \times 1}$, $(k = 0, \ldots, q-1)$, where $\lambda_c$ is the wavelength and $d = \lambda_c/2$ is the inter-element distance of the ULA, $\boldsymbol{s} \in \mathbb{R}^{q \times 1}$ is the source data, $\boldsymbol{n} \in \mathbb{C}^{m \times 1}$ is the white Gaussian noise, and $(\cdot)^T$ stands for the transpose. The output of a narrowband beamformer is

$$y = \boldsymbol{w}^H \boldsymbol{r}, \tag{2}$$

where $\boldsymbol{w} = [w_1, \ldots, w_m]^T \in \mathbb{C}^{m \times 1}$ is the complex weight vector of the adaptive filter, and $(\cdot)^H$ stands for the Hermitian transpose.

The conventional LCMV optimization problem determines the filter parameters by solving

$$
\begin{aligned}
\text{minimize} \quad & \mathbb{E}[|y|^2] = \boldsymbol{w}^H \boldsymbol{R} \boldsymbol{w} \\
\text{subject to} \quad & \boldsymbol{w}^H \boldsymbol{a}(\theta_0) = \gamma,
\end{aligned}
\tag{3}
$$

where $\gamma$ is a constant corresponding to the magnitude of the constraint. The objective of (3) is to minimize the array output power while maintaining the contribution from $\theta_0$ constant. The weight solution is $\boldsymbol{w}_{\text{opt}} = \frac{\gamma \boldsymbol{R}^{-1} \boldsymbol{a}(\theta_0)}{\boldsymbol{a}^H(\theta_0) \boldsymbol{R}^{-1} \boldsymbol{a}(\theta_0)}$. Many adaptive filtering algorithms have been reported to solve optimization problems. Among them the CG-type algorithms are promising due to their attractive tradeoff between performance and complexity. In what follows, we consider a constrained adaptive filtering framework with the SM technique.

## III. Constrained SM Adaptive Filtering Framework

For conventional constrained adaptive filtering techniques, the received vector $\boldsymbol{r}$ is processed by the LCMV filter $\boldsymbol{w}$ to generate the array output $y$. The parameters of $\boldsymbol{w}$ have to be updated at each time instant, which results in a high computational load when the filter length (the number of elements in the array) is large. Differently from conventional algorithms, SM-based techniques employ a check block with a specified bound $\delta$ on the amplitude of the array output $y$. The bound is assumed here to be time-varying and related to the previous estimated parameters of the weight vector. It is clear that, at each time instant, some valid estimates of $\boldsymbol{w}$ are consistent with the bound. Therefore, the solution to the new strategy is a set in the parameter space [28]. Some estimates even satisfy the constrained condition with respect to different $\boldsymbol{r}$ for different time instants. The parameter update is performed only if the bounded constraint $|y|^2 \leq \delta^2$ cannot be satisfied. This strategy leads to data-selective updates and ensures that all the updated parameters satisfy the constraint for the current instant. It implies that this scheme will have a reduced computational complexity.

Let $\mathcal{H}_i$ denote the set containing all the estimates of $\boldsymbol{w}$ for which the associated array output at time instant "$i$" is upper bounded in magnitude by $\delta$, which is

$$\mathcal{H}_i = \left\{ \boldsymbol{w} \in \mathbb{C}^{m \times 1} : \ |y|^2 \leq \delta^2 \right\}, \tag{4}$$

where $\mathcal{H}_i$ is bounded by a set of hyperplanes that correspond to the estimates of $\boldsymbol{w}$. It is given to enforce the bounded constraint and thus is referred to as the *constraint set*. It should be remarked that $\mathcal{H}_i$ includes the estimates for the current instant. We then define the exact *feasibility set* $\Theta_i$ as the intersection of the constraint sets over the time instants $l = 1, \ldots, i$, which is described by

$$\Theta_i = \bigcap_{\substack{l=1 \\ (s_0, \boldsymbol{r}) \in \boldsymbol{S}}}^{i} \mathcal{H}_l, \tag{5}$$

where $s_0$ corresponds to the transmitted data of the desired user and $\boldsymbol{S}$ is the set including all possible data pairs $\{s_0, \boldsymbol{r}\}$. The constrained SM approach only performs updates for those estimates which belong to this feasibility set. In practice, it is impossible to traverse all possible data pairs. Under this condition, the constrained SM framework works with the *membership set* constructed from the observed data pairs, which is given by $\Psi_i = \bigcap_{l=1}^{i} \mathcal{H}_l$. Note that the feasibility set is included in the membership set. The two sets are equal only if all possible data pairs are traversed up to time instant $i$.

The constrained adaptive filtering strategy that will be detailed in what follows employs the SM technique with a constrained adaptive CG-based algorithm. The convergence and tracking performance is enhanced due to the variable forgetting factor, whereas the computational complexity is reduced due to the data-selective updates which bring the update rate down. Note that, according to the nature of many dynamic scenarios, the time-varying bound should be selected appropriately to describe the characteristics of the environment and to improve the output performance. We will detail the development of the algorithm in the next section.

## IV. Proposed SM-CG Adaptive Algorithm

In this section, we introduce a constrained optimization strategy that combines the SM technique with the LCMV design and utilizes the CG-based adaptive filtering algorithm, as depicted in Fig. 1. The basic idea is to introduce a SM check that leads to the update of the beamformer when $|y|^2 > \delta^2$ and no update when $|y|^2 \le \delta^2$.

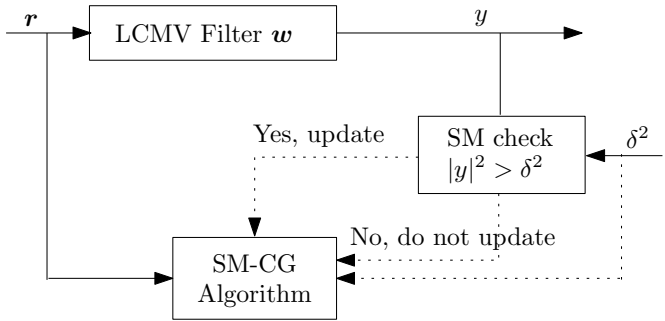

Fig. 1. Proposed constrained SM adaptive filtering strategy.

We develop a CG-based adaptive algorithm based on the SM technique to compute the parameters of the adaptive beamformer. The development starts from an LCMV optimization problem that incorporates the time-varying bound into the set of constraints as given by

$$\begin{aligned}
&\text{minimize } \boldsymbol{w}^H \hat{\boldsymbol{R}} \boldsymbol{w} \\
&\text{subject to } \boldsymbol{w}^H \boldsymbol{a}(\theta_0) = \gamma \text{ whenever } |y|^2 > \delta^2,
\end{aligned} \tag{6}$$

where $\delta$ determines a set of estimates of $\boldsymbol{w}$ within the constraint set $\mathcal{H}$ for each time instant and $\hat{\boldsymbol{R}}$ is an estimate of the covariance matrix of the received vector $\boldsymbol{r}$ described by $\boldsymbol{R} = E[\boldsymbol{r}\boldsymbol{r}^H]$.

We transform the constrained optimization problem into an unconstrained one using the method of Lagrange multipliers [1] and consider an equality as the second constraint in (6). The algorithm will update whenever $|y|^2 > \delta^2$. We then employ an exponentially weighted cost function with time index $i$ expressed by

$$\begin{aligned}
L(\boldsymbol{w}(i), \lambda_1(i), \lambda_2) = &\sum_{l=1}^{i-1} \lambda_1^{i-l} \boldsymbol{w}^H(i)\boldsymbol{r}(l)\boldsymbol{r}^H(l)\boldsymbol{w}(i) \\
&+ 2\lambda_1(i)\Re\{|y(i)|^2 - \delta^2(i)\} + 2\lambda_2\Re\{\boldsymbol{w}^H(i)\boldsymbol{a}(\theta_0) - \gamma\},
\end{aligned} \tag{7}$$

where $\Re\{\cdot\}$ selects the real part of the argument, $\lambda_1(i)$ plays the role of forgetting factor and Lagrange multiplier with respect to the bounded constraint and is calculated only if the bounded constraint is satisfied. The scalar $\lambda_2$ is another Lagrange multiplier that ensures the constraint on the array response of the desired user.

For further derivations, we assume that $\lambda_1(i)$ is close to 1, which is consistent with the setting of the forgetting factor in [1]. Under this condition, the solution of the beamforming problem in (6) is given by $\boldsymbol{w}(i) = \frac{\gamma \hat{\boldsymbol{R}}^{-1}(i)\boldsymbol{a}(\theta_0)}{\boldsymbol{a}^H(\theta_0)\hat{\boldsymbol{R}}^{-1}(i)\boldsymbol{a}(\theta_0)}$, where $\hat{\boldsymbol{R}}(i) = \hat{\boldsymbol{R}}(i-1) + \lambda_1(i)\boldsymbol{r}(i)\boldsymbol{r}^H(i)$ is an alternative form to estimate the covariance matrix $\boldsymbol{R}$. This expression is similar to that of the conventional LCMV solution. The main difference is that, in the proposed scheme, the bounded constraint reduces the update rate and provides valid estimates of $\boldsymbol{w}$ to construct the feasibility set $\Psi_i$.

To solve the optimization problem, we define a CG-based vector that exploits a relation with the received covariance matrix and the array response of the desired user given by

$$\boldsymbol{v}(i) = \hat{\boldsymbol{R}}^{-1}(i)\boldsymbol{a}(\theta_0). \tag{8}$$

Using this relation, the weight vector can be written as

$$\boldsymbol{w}(i) = \frac{\gamma \boldsymbol{v}(i)}{\boldsymbol{a}^H(\theta_0)\boldsymbol{v}(i)}, \tag{9}$$

where $\boldsymbol{v}(i)$ is regarded as an intermediate parameter vector to guarantee the constraints and to avoid the matrix inverse. We intend to perform one iteration per update to calculate $\boldsymbol{v}(i)$ and the weight vector $\boldsymbol{w}(i)$.

According to the methodology of the CG algorithm [11], the CG-based update rule of the auxiliary vector $\boldsymbol{v}(i)$ is given by

$$\boldsymbol{v}(i) = \boldsymbol{v}(i-1) + \alpha(i)\boldsymbol{p}(i), \tag{10}$$

where $\alpha(i)$ is a step-size coefficient and $\boldsymbol{p}(i)$ is a direction vector that is obtained by a linear combination of the previous direction vector and a negative gradient vector $\boldsymbol{g}(i)$ of $J(\boldsymbol{v}(i)) = \boldsymbol{v}^H(i)\hat{\boldsymbol{R}}(i)\boldsymbol{v}(i) - 2\Re\{\boldsymbol{v}^H(i)\boldsymbol{a}(\theta_0)\}$ with respect to $\boldsymbol{v}^*(i)$, which is

$$\begin{aligned}
\boldsymbol{g}(i) &= \boldsymbol{a}(\theta_0) - \hat{\boldsymbol{R}}(i)\boldsymbol{v}(i) \\
&= \boldsymbol{g}(i-1) - \alpha(i)\hat{\boldsymbol{R}}(i)\boldsymbol{p}(i) - \lambda_1(i)\boldsymbol{r}(i)\boldsymbol{r}^H(i)\boldsymbol{v}(i-1).
\end{aligned} \tag{11}$$

We employ an inexact line search scheme with reduced complexity to calculate $\boldsymbol{v}(i)$. The step-size coefficient should satisfy the convergence bound $0 \le \boldsymbol{p}^H(i)\boldsymbol{g}(i) \le 0.5\boldsymbol{p}^H(i)\boldsymbol{g}(i-1)$ according to [11]. From this bound and (11), we have

$$\begin{aligned}
&\frac{0.5\boldsymbol{p}^H(i)\boldsymbol{g}(i-1) - \lambda_1(i)\boldsymbol{p}^H(i)\boldsymbol{r}(i)\boldsymbol{r}^H(i)\boldsymbol{v}(i-1)}{\boldsymbol{p}^H(i)\hat{\boldsymbol{R}}(i)\boldsymbol{p}(i)} \le \\
&\alpha(i) \le \frac{\boldsymbol{p}^H(i)\boldsymbol{g}(i-1) - \lambda_1(i)\boldsymbol{p}^H(i)\boldsymbol{r}(i)\boldsymbol{r}^H(i)\boldsymbol{v}(i-1)}{\boldsymbol{p}^H(i)\hat{\boldsymbol{R}}(i)\boldsymbol{p}(i)}.
\end{aligned} \tag{12}$$

The relations in (12) are satisfied if $\alpha(i)$ is

$$\alpha(i) = \frac{(1-\eta)\boldsymbol{p}^H(i)\boldsymbol{g}(i-1) - \lambda_1(i)\boldsymbol{p}^H(i)\boldsymbol{r}(i)\boldsymbol{r}^H(i)\boldsymbol{v}(i-1)}{\boldsymbol{p}^H(i)\hat{\boldsymbol{R}}(i)\boldsymbol{p}(i)}, \tag{13}$$

where $0 \le \eta \le 0.5$.

Substituting (10) and (13) into the bounded constraint, and performing some algebraic manipulations, we obtain the expression for the coefficient $\lambda_1(i)$, which is

$$\lambda_1(i) = \frac{\lambda_{11}(i) - \lambda_{12}(i)}{\lambda_{13}(i) - \lambda_{14}(i)}, \tag{14}$$

where

$\lambda_{11}(i) = \tau_1(i)\operatorname{sign}\{\tau_1(i) - \tau_2(i)\}$
$\lambda_{12}(i) = \tau_3(i)\operatorname{sign}\{\tau_3(i) - \tau_4(i)\}$
$\lambda_{13}(i) = \tau_2(i)\operatorname{sign}\{\tau_1(i) - \tau_2(i)\}$
$\lambda_{14}(i) = \tau_4(i)\operatorname{sign}\{\tau_3(i) - \tau_4(i)\}$
$\tau_1(i) = \delta(i)\boldsymbol{v}^H(i-1)\boldsymbol{a}(\theta_0)\boldsymbol{p}^H(i)\hat{\boldsymbol{R}}(i)\boldsymbol{p}(i) + \delta(i)(1-\eta)\boldsymbol{g}^H(i-1)\boldsymbol{p}(i)\boldsymbol{p}^H(i)\boldsymbol{a}(\theta_0)$
$\tau_2(i) = \boldsymbol{v}^H(i-1)\boldsymbol{r}(i)\boldsymbol{r}^H(i)\boldsymbol{p}(i)\boldsymbol{p}^H(i)\boldsymbol{a}(\theta_0)$
$\tau_3(i) = \boldsymbol{v}^H(i-1)\boldsymbol{r}(i)\boldsymbol{p}^H(i)\hat{\boldsymbol{R}}(i)\boldsymbol{p}(i) + (1-\eta)\boldsymbol{g}^H(i-1)\boldsymbol{p}(i)\boldsymbol{p}^H(i)\boldsymbol{r}(i)$
$\tau_4(i) = \boldsymbol{v}^H(i-1)\boldsymbol{r}(i)\boldsymbol{r}^H(i)\boldsymbol{p}(i)\boldsymbol{p}^H(i)\boldsymbol{r}(i)$.

The details of the derivation are shown in the appendix.

It should be remarked that $\lambda_1(i)$ is adaptive, and involves the estimated parameters and the time-varying bound. Unlike in conventional adaptive algorithms that employ a forgetting factor which is usually a fixed real value close to 1, in the proposed SM-CG algorithm $\lambda_1(i)$ is adaptive, can follow the changes of the scenarios and improves the tracking performance. This is a feature that is not available in the algorithm in [31]. Furthermore, it avoids the divergence problem caused by the setting of the forgetting factor and thus ensures a stable steady-state performance of the proposed algorithm.

The direction vector $\boldsymbol{p}(i)$ is

$$\boldsymbol{p}(i+1) = \boldsymbol{g}(i) + \beta(i)\boldsymbol{p}(i), \tag{15}$$

where $\beta(i)$ is chosen to provide conjugacy [34] for the direction vector and is given by

$$\beta(i) = -\frac{\boldsymbol{p}^H(i)\hat{\boldsymbol{R}}(i)\boldsymbol{g}(i)}{\boldsymbol{p}^H(i)\hat{\boldsymbol{R}}(i)\boldsymbol{p}(i)}. \tag{16}$$

The proposed SM-CG algorithm is summarized in Table I, where the initialization is set to enforce the constraint with respect to $\boldsymbol{a}(\theta_0)$ and to start the update. The coefficient $\lambda_1(i)$ is calculated only if the bounded constraint cannot be satisfied, so as the update of the parameters. During the update, the proposed algorithm runs only one iteration to calculate the CG-based vector for the weight solution, which requires much less computational cost in comparison with existing CG-based algorithms [11], [34].

## V. Schemes to Compute Time-Varying Bounds

The bound $\delta$ is a scalar coefficient for the SM technique to check if the parameter update should be performed or not. It is an important parameter to measure the quality of the estimates that could be included in the feasibility set $\Theta_i$. In [25], [28]–[32], several predetermined bounding schemes have been reported for development of the adaptive algorithms, which achieve reduced complexity without performance degradation. However, they may suffer from the risk of underbounding (the bound is smaller than the actual one) or overbounding (the bound is larger than the actual one). Besides, the

### TABLE I
### THE PROPOSED SM-CG ALGORITHM

Initialization:
$\boldsymbol{g}(0) = \boldsymbol{p}(1) = \boldsymbol{a}(\theta_0);\quad \boldsymbol{w}(0) = \boldsymbol{a}(\theta_0)/\|\boldsymbol{a}(\theta_0)\|^2.$
**For each time instant** $i = 1, \ldots, N$
$\quad y(i) = \boldsymbol{w}^H(i-1)\boldsymbol{r}(i)$
$\quad \delta(i)$ (PDB or PIDB scheme) in (17) or (23)
$\quad$**if** $\quad |y(i)|^2 \geq \delta^2(i)$
$\quad\quad \lambda_1(i) = \frac{\lambda_{11}(i) - \lambda_{12}(i)}{\lambda_{13}(i) - \lambda_{14}(i)}$
$\quad\quad \hat{\boldsymbol{R}}(i) = \hat{\boldsymbol{R}}(i-1) + \lambda_1(i)\boldsymbol{r}(i)\boldsymbol{r}^H(i)$
$\quad\quad \alpha(i) = \frac{(1-\eta)\boldsymbol{p}^H(i)\boldsymbol{g}(i-1) - \lambda_1(i)\boldsymbol{p}^H(i)\boldsymbol{r}(i)\boldsymbol{r}^H(i)\boldsymbol{v}(i-1)}{\boldsymbol{p}^H(i)\hat{\boldsymbol{R}}(i)\boldsymbol{p}(i)}$
$\quad\quad \boldsymbol{v}(i) = \boldsymbol{v}(i-1) + \alpha(i)\boldsymbol{p}(i)$
$\quad\quad \boldsymbol{g}(i) = \boldsymbol{g}(i-1) - \alpha(i)\hat{\boldsymbol{R}}(i)\boldsymbol{p}(i) - \lambda_1(i)\boldsymbol{r}(i)\boldsymbol{r}^H(i)\boldsymbol{v}(i-1)$
$\quad\quad \beta(i) = -\frac{\boldsymbol{p}^H(i)\hat{\boldsymbol{R}}(i)\boldsymbol{g}(i)}{\boldsymbol{p}^H(i)\hat{\boldsymbol{R}}(i)\boldsymbol{p}(i)}$
$\quad\quad \boldsymbol{p}(i+1) = \boldsymbol{g}(i) + \beta(i)\boldsymbol{p}(i)$
$\quad$**else**
$\quad\quad \boldsymbol{v}(i) = \boldsymbol{v}(i-1)$
$\quad$**end**
$\quad \boldsymbol{w}(i) = \frac{\gamma\boldsymbol{v}(i)}{\boldsymbol{a}^H(\theta_0)\boldsymbol{v}(i)}$

predetermined bound cannot reflect the characteristics (time-varying nature) of the environment and thus may result in poor convergence and tracking performance in dynamic scenarios.

Here we introduce two alternative bounding schemes that are time-varying instead of the predetermined bounds to enhance the convergence and tracking performances. The first scheme, which is called parameter dependent bound (PDB), is an extension to the work reported in [35]–[37] which involves the CG-based parameters of the proposed SM-CG algorithm. The PDB recursion is

$$\delta(i) = \varrho\delta(i-1) + (1-\varrho)\sqrt{\varsigma\|\boldsymbol{w}\|^2\hat{\sigma}_n^2(i)}, \tag{17}$$

where $\varrho$ is a forgetting factor that should be adjusted to guarantee a proper time-averaged estimate of $\boldsymbol{w}(i)$, $\varsigma$ ($\varsigma > 1$) is a tuning coefficient to adjust $\|\boldsymbol{w}(i)\|^2\hat{\sigma}_n^2(i)$ in order to provide information on the evolution of the power of $\boldsymbol{w}(i)$, and $\hat{\sigma}_n^2(i)$ is an estimate of the noise power. We assume that the noise power is known beforehand at the receiver. This time-varying bound formulates a relation between the estimated parameters and the environmental coefficients. It provides a smoother evolution of the weight vector trajectory and thus avoids too high or low values of the squared norm of the weight vector.

Another time-varying bounding scheme combines the PDB with an interference estimation procedure, which is a little more complex than the PDB but has an improved performance since it provides more information about the environment for parameter estimation [39]. It is termed parameter and interference dependent bound (PIDB) and is an extension of [39] for array signal processing. Differently from [39], this scheme employs the steering vector of the desired user and the beamforming filter to subtract the components of the desired user from the output of the steering vector for estimating the multiple access interference (MAI) power level, as depicted in Fig. 2.

In this structure, the received vector $\boldsymbol{r}(i)$ is processed by the SM-CG strategy and the steering vector $\boldsymbol{a}(\theta_0)$, respectively, to generate the array output $y(i)$ and the component with respect

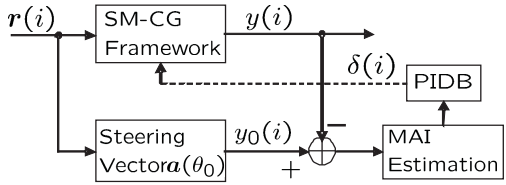

Fig. 2. PIDB bound structure

to the desired user $y_0(i)$, which is

$$
\begin{aligned}
y_0(i) &= \boldsymbol{a}^H(\theta_0)\boldsymbol{r}(i) \\
&= s_0(i) + \underbrace{\sum_{k=1}^{q-1}\boldsymbol{a}^H(\theta_0)\boldsymbol{a}(\theta_k)s_k(i)}_{\text{MAI}} + \underbrace{\boldsymbol{a}^H(\theta_0)\boldsymbol{n}(i)}_{\text{noise}}, \quad (18)
\end{aligned}
$$

where $s_k(i)$ $(k = 0, \ldots, q-1)$ denotes the transmitted data from the $k$th user.

The second-order statistics of the output of the steering vector is

$$
\begin{aligned}
\mathbb{E}[|y_0(i)|^2] &= \sum_{l=1}^{q-1}\sum_{k=1}^{q-1}\mathbb{E}[s_l(i)s_k^*(i)]\boldsymbol{a}^H(\theta_0)\boldsymbol{a}(\theta_l)\boldsymbol{a}^H(\theta_k)\boldsymbol{a}(\theta_0) \\
&\quad + \mathbb{E}[|s_0(i)|^2] + \boldsymbol{a}^H(\theta_0)\mathbb{E}[\boldsymbol{n}(i)\boldsymbol{n}^H(i)]\boldsymbol{a}(\theta_0) \\
&= \sum_{k=1}^{q-1}\mathbb{E}[|s_k(i)|^2]\boldsymbol{a}^H(\theta_0)\boldsymbol{a}(\theta_k)\boldsymbol{a}^H(\theta_k)\boldsymbol{a}(\theta_0) \\
&\quad + \mathbb{E}[|s_0(i)|^2] + \sigma_n^2, \\
&\hspace{6cm} (19)
\end{aligned}
$$

where $\sigma_n^2$ is the noise variance and the second expression is obtained since we assume that source data are transmitted independently. From (19), the second-order statistics can be used to identify the sum of the power level of the MAI and noise terms.

Assuming that the algorithm converges, we should have $y(i) \to s_0(i)$. According to the PIDB structure, the instantaneous estimate of the MAI and noise can be obtained by subtracting $y(i)$ from $y_0(i)$, which is given by

$$
\begin{aligned}
e_0(i) &= y_0(i) - y(i) \\
&\approx \underbrace{\sum_{k=1}^{q-1}\mathbb{E}[|s_k(i)|^2]\boldsymbol{a}^H(\theta_0)\boldsymbol{a}(\theta_k)s_k(i)}_{\text{MAI}} + \underbrace{\boldsymbol{a}^H(\theta_0)\boldsymbol{n}(i)}_{\text{noise}}. \\
&\hspace{6cm} (20)
\end{aligned}
$$

Taking the expectation with respect to $|e_0(i)|^2$ and considering the MAI and noise uncorrelated, we have

$$
\mathbb{E}[|e_0(i)|^2] = \sum_{k=1}^{q-1}\mathbb{E}[|s_k(i)|^2]\boldsymbol{a}^H(\theta_0)\boldsymbol{a}(\theta_k)\boldsymbol{a}^H(\theta_k)\boldsymbol{a}(\theta_0) + \sigma_n^2, \quad (21)
$$

where it represents the interference and noise power. By using time averages of the instantaneous values, we could obtain an estimate of (21), which is

$$
v(i) = \varrho v(i-1) + (1-\varrho)|e_0(i)|^2, \quad (22)
$$

where $\varrho$ is a forgetting factor to ensure a proper time-averaged estimate of the interference and noise power.

The information of the interference and noise power can be incorporated into the bounding scheme, which leads to the PIDB expression

$$
\delta(i) = \varrho\delta(i-1) + (1-\varrho)[\sqrt{\varepsilon v(i)} + \sqrt{\varsigma\|\boldsymbol{w}(i)\|^2\hat{\sigma}_n^2(i)}], \quad (23)
$$

where $\varepsilon$ is a weighting parameter that should be set. The expression in (22) avoids instantaneous values that are undesirably too high or too low, and thus circumvents inappropriate estimates of $\delta(i)$. Compared with (17), the PIDB involves the estimation of the interference and noise power and provides more information to track the characteristics of the environment, which benefits the convergence and tracking performance.

## VI. Complexity Analysis

In this section, we show the computational complexity requirements for the proposed SM-CG algorithm and compare it with those of the existing algorithms. The computational cost is measured in terms of the number of complex arithmetic operations, i.e., additions and multiplications. The results are listed in Table II, where $m$ is the number of sensor elements, $N$ is the total number of snapshots, $\tau$ $(0 < \tau \leq 1)$ is the update rate for the adaptive algorithms with the SM technique, and $L$ is the number of the observed received vector in the signal matrix of the AP and DS-CG algorithms. Note that the calculations of the time-varying bound are based on (17) for all the algorithms.

From Table II, the update rate $\tau$ impacts the complexity significantly. Specifically, for a small value of $\tau$, the complexity of the algorithms with the SM technique is much lower than their conventional counterparts that have $\tau = 100\%$ since the SM algorithms only perform updates for a small number of snapshots. For a very large $\tau$ (e.g., $\tau = 1$), the SM-based algorithms are a little more complex due to the calculations of the time-varying bound and the related coefficients. In most cases, the update rate has a relatively small value (around $10\% - 20\%$) and thus leads to a reduced complexity.

Fig. 3 provides a more direct way to illustrate the computational requirements for the adaptive algorithms. It includes the complexity in terms of additions and multiplications versus the number of sensor elements $m$. It should be remarked that $\tau$ is different with respect to the different algorithms, which are set to ensure a good output performance. The specific values are given in the figure. From Fig. 3, the adaptive algorithms with the SM technique require much less computational cost than their conventional counterparts. The proposed SM-CG algorithm keeps a small update rate ($\tau = 6.0\%$), which is lower than those of other SM-based algorithms. The curves of the SM-CG are situated between those of the SM-SG and the SM-RLS algorithms. As $m$ increases, the complexity of the proposed SM-CG algorithm is slightly higher than that of the SM-AP algorithm. However, the latter cannot achieve as good performance as that of the proposed one. This fact will be shown in the simulation results. We also compare the proposed SM-CG algorithm with the DS-CG algorithm of

## TABLE II
### Computational complexity of algorithms

| Algorithm | Additions | Multiplications |
|---|---|---|
| SG | $N(3m-1)$ | $N(4m+1)$ |
| SM-SG | $2Nm+3\tau Nm$ | $N(2m+5)+\tau N(4m+3)$ |
| RLS | $N(4m^2-m-1)$ | $N(5m^2+5m-1)$ |
| SM-RLS | $2Nm+\tau N(4m^2-1)$ | $N(2m+5)+\tau N(5m^2+6m+2)$ |
| SM-AP | $N(2m+1)+\tau N[(m-1)L^2+mL+1]$ | $N(2m+5)+\tau N[L^3+mL^2+(m+1)L+m+2]$ |
| CG | $N(2m^2+7m+1)$ | $N(2m^2+11m+5)$ |
| DS-CG | $\tau N(2m^2+8m-2)+LN(m-1)$ | $\tau N(2m^2+9m+3)+LNm$ |
| SM-CG | $2Nm+\tau N(2m^2+8m+6)$ | $N(2m+5)+\tau N(2m^2+9m+22)$ |

[31]. As the update rate of the DS-CG algorithm (22.1%) is much higher than that of the proposed algorithm (6.0%), the complexity of the former is typically higher than the latter. The number of the received vectors $L$ ($L = 3$ in this experiment) used for processing also increases the complexity.

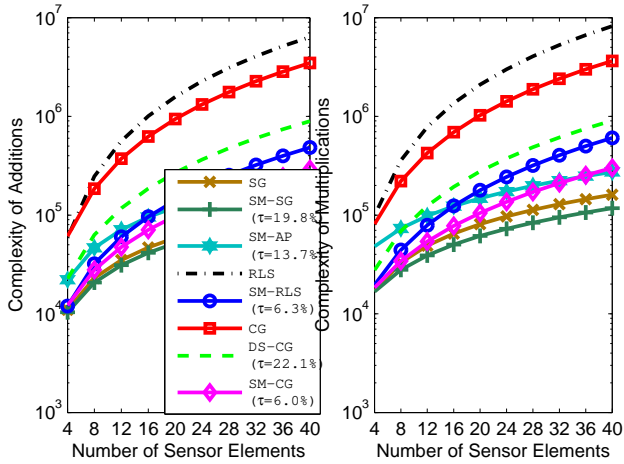

Fig. 3. Complexity in terms of arithmetic operations versus the number of sensor elements $m$.

## VII. Simulations

We evaluate the SINR performance of the proposed SM-CG algorithm with the LCMV design and compare it with the existing SG, RLS, AP, and CG algorithms with/without the SM technique [1], [11]–[26], [28], [32]. In all simulations, we assume that there is one desired user in the system and the related DOA is known by the receiver. Simulations are performed with a ULA containing $m = 16$ sensor elements with half-wavelength interelement spacing. The results are averaged by 500 runs. We consider the binary phase shift keying (BPSK) modulation scheme and set $\gamma = 1$.

In Fig. 4, we first compare the SINR performance of the proposed SM-CG algorithm with the existing DS-CG algorithm in [31]. We consider a scenario with $q = 10$ users in the system. The input signal-to-noise ratio (SNR) is 10 dB and the interference-to-noise ratio (INR) is 30 dB. The total number of snapshots is $N = 4000$. The coefficients are $\lambda = 0.999$, $\eta = 0.8$, $L = 3$, $\delta = 0.001$ for the DS-CG algorithm. The bound used for the analyzed algorithms is fixed and chosen as $\sqrt{5\sigma_n^2}$. The curves of the RLS algorithm and the MVDR solution are included for comparison. It is clear that the convergence rate of the proposed SM-CG algorithm is faster than that of the DS-CG algorithm. The proposed SM-CG algorithm employs the variable forgetting factor to enhance the performance. The update rate of the proposed algorithm (6.0%) is much lower than that of the DS-CG algorithm (22.1%). It should be remarked that we have compared the SM-CG and DS-CG algorithms in numerous other scenarios. The SM-CG algorithm has consistently performed better than the DS-CG algorithm and required lower update rates than those of the DS-CG technique.

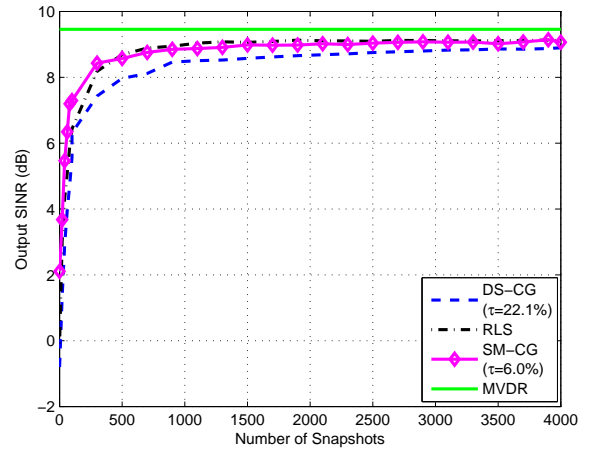

Fig. 4. Output SINR versus the number of snapshots.

In Fig. 5, we show the SINR performance of the proposed SM-CG algorithm with the fixed and time-varying bounds. There are $q = 10$ users in the system. The input signal-to-noise ratio (SNR) is 10 dB and the interference-to-noise ratio (INR) is 35 dB. The total number of snapshots is $N = 3000$. The proposed algorithm is performed with the fixed bounds $\delta = 0.8, 1.0, 1.3$, the PDB and PIDB schemes. The coefficients are $\varsigma = 21$, $\varrho = 0.9$, $\eta = 0.5$ for the PDB, and $\varsigma = 0.98$, $\varrho = 19$, $\eta = 0.5$, $\varepsilon = 0.001$ for the PIDB, respectively. We find that the curves with the fixed bounds experience an increase trend and then a decrease trend due to the divergence problem [1]. In these cases, it is necessary to utilize the regularization procedure periodically and set an appropriate forgetting factor to avoid the divergence. For the proposed SM-CG algorithm, this problem is minimized since the coefficient $\lambda_1(i)$ is variable according to the time-varying bound and thus circumvents the risk of divergence. The curves with the time-varying bounds have much lower update rates than those with the fixed bounds. From Fig. 5, the curve with the PIDB exhibits a slightly superior convergence performance over that with the PDB since the former involves the estimation of the interference and noise power. For comparison and clear statement, we only show the results of the proposed and existing algorithms equipped with the PIDB scheme in what follows.

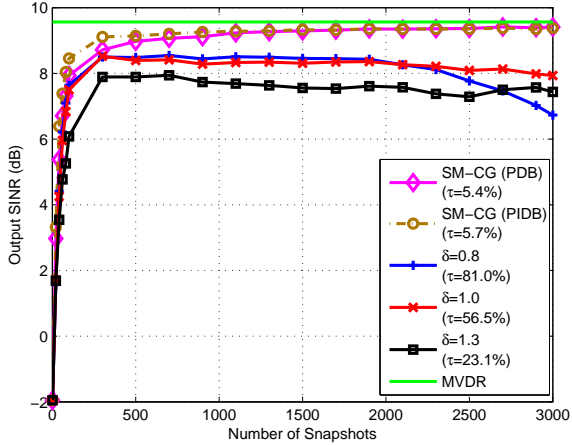

Fig. 5. Output SINR versus the number of snapshots.

In Fig. 6, we show the output SINR performance of the proposed and existing algorithms versus the number of snapshots. The scenario is the same as that in Fig. 5 except the input INR= 30 dB. The related coefficients are $\varsigma = 21$, $\varrho = 0.9$, and $\eta = 0.5$. The coefficient $\lambda_1(i)$ is limited to $0.1 \leq \lambda_1(i) \leq 0.999$ in accordance with the setting of the forgetting factor. From Fig. 6, the algorithms with the SM technique show superior convergence over the standard algorithms. The proposed SM-CG algorithm has a fast convergence that is close to that of the SM-RLS algorithm and reaches the high steady-state performance close to the minimum variance distortionless response (MVDR) solution. It should be remarked that the update rate for the proposed SM-CG algorithm is only $\tau = 6.0\%$ (178 updates for 3000 snapshots), which is lower than those of the existing algorithms and reduces the computational cost significantly.

The next experiment exhibits the performance of the time-varying bound $\delta(i)$ for the proposed and existing SM-based algorithms under the same scenario as in Fig. 6. The time-varying bound reflects the convergence performance and also the update rate. From Fig. 7, we find that the curve of the time-varying bound for the proposed SM-CG algorithm is close to that for the SM-RLS algorithm. It reaches the steady-state rapidly and has a small update rate.

Fig. 8 evaluates the SINR performance of the algorithms with different input SNR values. There are $q = 10$ users in the system. The input SINR varies between 0 dB and 30 dB. It is clear that the SINR values of all the algorithms increases with the increase of the SNR values, as expected. The proposed SM-CG algorithm exhibits a performance that is close to the SM-RLS method and better than other existing techniques. In Table III, we provide the update rates for the proposed and existing algorithms with respect to different input SNR values. It shows that the update rates of the proposed SM-CG algorithm always keep a low level with different SNR values.

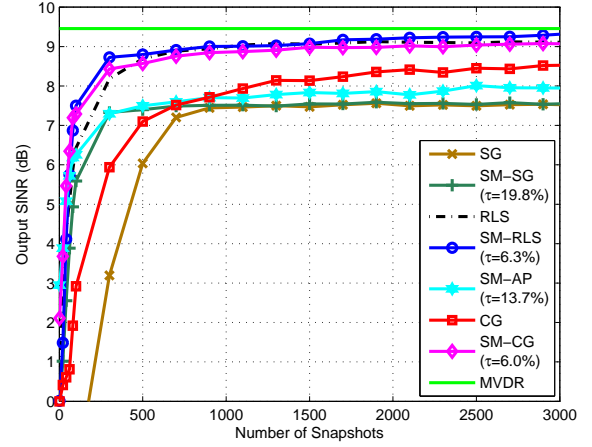

Fig. 6. Output SINR versus the number of snapshots.

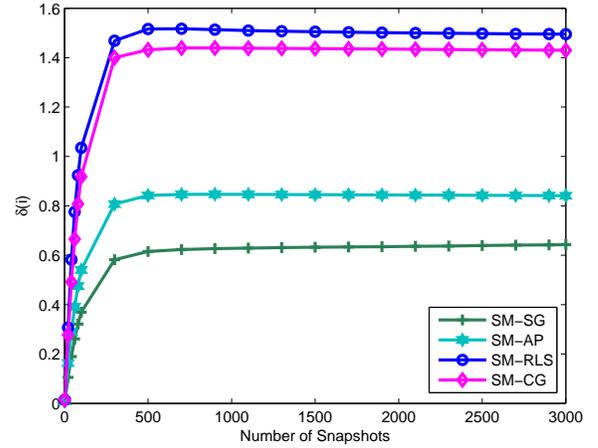

Fig. 7. Time-varying bound versus the number of snapshots.

The last experiment is provided to show the tracking performance of the proposed and existing algorithms in a non-stationary scenario, namely, when the number of interferers changes in the system. The system starts with $q = 8$ users. The input SNR is 10 dB and the INR is 35 dB. The proposed algorithm converges rapidly to the steady-state. The system experiences a sudden change at $N = 3000$. We have 4 more interferers entering the system. This change degrades the SINR performance for all the algorithms. The algorithms with the SM technique track this change quickly and reach the steady-state again since the data-selective updates reduce the number of updates and thus lead to a faster convergence. Also, the PDB scheme provides information for the algorithms to follow the changes of the scenario. The proposed SM-CG algorithm exhibits a good tracking performance with a relatively low update rate ($\tau = 6.2\%$).

TABLE III
Update Rates For Different Algorithms

|         | 0 (dB) | 5 (dB) | 10 (dB) | 15 (dB) | 20 (dB) | 25 (dB) | 30 (dB) |
|---------|--------|--------|---------|---------|---------|---------|---------|
| SM-SG   | 14.7%  | 16.9%  | 19.8%   | 21.2%   | 15.5%   | 16.8%   | 17.2%   |
| SM-RLS  | 6.8%   | 7.2%   | 6.3%    | 7.8%    | 10.2%   | 8.8%    | 8.3%    |
| SM-AP   | 9.3%   | 11.4%  | 13.7%   | 14.7%   | 13.1%   | 13.8%   | 10.7%   |
| SM-CG   | 5.6%   | 5.9%   | 6.0%    | 6.3%    | 6.2%    | 6.0%    | 5.9%    |

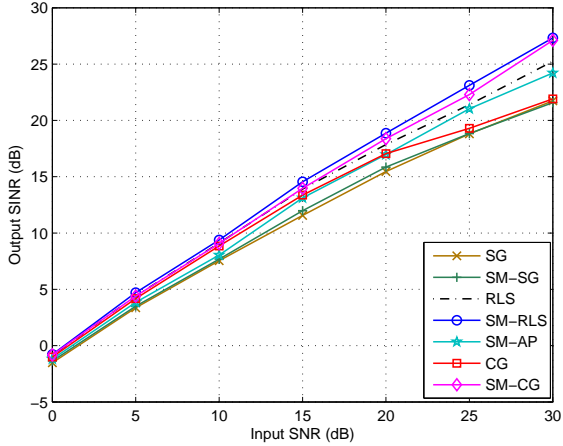

Fig. 8. Output SINR versus different input SNR values.

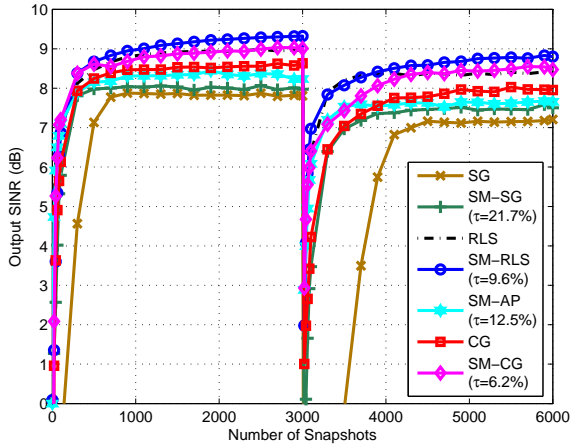

Fig. 9. Output SINR versus the number of snapshots in dynamic scenario with additional users enter and/or leave the system.

## VIII. Conclusions

We have introduced a new adaptive filtering strategy that combines the SM technique with the adaptive CG algorithm for the design of LCMV beamformers. A bounded constraint has been incorporated in this strategy to realize the data-selective updates for the weight solution with a reduced computational cost. During the update, a CG-based parameter vector is used to exploit a relation between the covariance matrix inverse and the steering vector of the desired user. The proposed strategy enforces the constraint with respect to the steering vector and avoids the matrix inversion and employs a time-varying forgetting factor. The proposed SM-CG algorithm constructs a space of feasible solutions that retain the contribution of the desired user and satisfy the bounded constraint. A complexity analysis has been carried out to show the computation reduction achieved by the proposed SM-CG algorithm. Simulation results have shown excellent convergence and tracking performance for the SM-CG algorithm.

## Derivation of (14)

In this appendix, we derive the expression of the coefficient $\lambda_1(i)$ in (14). Using the bounded constraint in (6) and $y(i) = \boldsymbol{w}^H(i)\boldsymbol{r}(i)$, we have

$$|\boldsymbol{w}^H(i)\boldsymbol{r}(i)|^2 = \delta^2(i). \tag{24}$$

Substituting the weight expression (9) into (24), the bounded constraint can be given by

$$|\boldsymbol{v}^H(i)\boldsymbol{r}(i)|^2 = \delta^2(i)|\boldsymbol{v}^H(i)\boldsymbol{a}(\theta_0)|^2, \tag{25}$$

which is equivalent to

$$\boldsymbol{v}^H(i)\boldsymbol{r}(i)\text{sign}\{\boldsymbol{v}^H(i)\boldsymbol{r}(i)\} = \delta(i)\boldsymbol{v}^H(i)\boldsymbol{a}(\theta_0)\text{sign}\{\boldsymbol{v}^H(i)\boldsymbol{a}(\theta_0)\}. \tag{26}$$

Substituting (13) into (10), the CG-based vector $\boldsymbol{v}(i)$ can be expressed by $\boldsymbol{v}(i-1)$, $\boldsymbol{p}(i)$, $\boldsymbol{g}(i-1)$, $\lambda_1(i)$, and $\delta(i)$, which is

$$\boldsymbol{v}(i) = \boldsymbol{v}(i-1)$$
$$+ \frac{\lambda_1(i)\boldsymbol{p}^H(i)[\boldsymbol{g}(i-1) - \boldsymbol{a}(\theta_0)]\boldsymbol{p}(i) - \eta\boldsymbol{p}^H(i)\boldsymbol{g}(i-1)\boldsymbol{p}(i)}{\boldsymbol{p}^H(i)\hat{\boldsymbol{R}}(i)\boldsymbol{p}(i)}. \tag{27}$$

Substituting (27) into (26), multiplying the term $\boldsymbol{p}^H(i)\hat{\boldsymbol{R}}(i)\boldsymbol{p}(i)$ on both sides of (26), and performing mathematical transformations, we obtain the expression for the coefficient $\lambda_1(i)$ in (14).